\documentclass[11pt]{article}

\usepackage[a4paper,margin=1in]{geometry}
\usepackage{amsmath,amssymb,amsthm}
\usepackage{hyperref}
\usepackage{enumitem}
\usepackage{booktabs}
\usepackage{xcolor}
\usepackage{url}
\usepackage{graphicx}

\hypersetup{
  colorlinks=true,
  linkcolor=blue!70!black,
  citecolor=green!50!black,
  urlcolor=blue!70!black
}

\title{Bisynchronous FIFOs and the FITO Category Mistake:\\
Silicon-Proven Interaction Primitives\\
for Distributed Coordination}

\author{
Paul Borrill \\
D{\AE}D{\AE}LUS \\
\texttt{paul@daedaelus.com}
}

\date{March 2026}

\begin{document}

\maketitle

\begin{abstract}
Bisynchronous FIFOs---hardware buffers that mediate data transfer
between independent clock domains without a shared global
timebase---have been designed, formally verified, and commercially
deployed in silicon for over four decades.  We survey this literature
from Chapiro's 1984 GALS thesis through Cummings's Gray-code pointer
techniques, Chelcea and Nowick's mixed-timing interfaces, Greenstreet's
STARI protocol, and the 2015 NVIDIA pausible bisynchronous FIFO, and
argue that this body of work constitutes a silicon-proven existence
proof against the Forward-In-Time-Only (FITO) assumption that pervades
distributed systems.  The central claim is that interaction-based
synchronization primitives---handshakes, mutual exclusion, and causal
flow control---can replace timestamp-based coordination at the most
demanding levels of digital engineering, directly undermining the FITO
assumption in protocols such as PTP, TSN, and conventional Ethernet.
We draw a structural parallel between on-chip bisynchronous
coordination and the Open Atomic Ethernet (OAE) architecture, and
identify the handshake---not the timestamp---as the fundamental
primitive for coordination between independent causal domains.
\end{abstract}

\tableofcontents
\newpage

\section{Introduction: The Synchronization Problem}
\label{sec:intro}

Every digital system that encompasses more than one clock domain faces a
fundamental question: how do you move data safely from a producer clocked
at one rate (or phase) to a consumer clocked at another?  The naive answer
is to sample the incoming signal with the receiver's clock, but this
invites metastability---a condition in which a flip-flop enters an
indeterminate state because the input violated setup or hold
timing.  Chaney and Molnar's 1973 report on anomalous synchronizer
behaviour~\cite{chaney1973} first documented that arbiters driven
under ambiguous timing can oscillate indefinitely, and Ginosar's
later tutorial~\cite{ginosar2011} consolidates the modern
understanding that metastability is ubiquitous and managed
probabilistically rather than eliminated.  The problem is not merely
theoretical.
Metastability failures have caused real-world system crashes, and the
literature treats the topic with a seriousness commensurate with the
stakes.  Indeed, the practical consequences of metastability and
transmission-line signalling anomalies were central concerns in the
design of the IEEE~P896 Futurebus standard~\cite{borrill1986futurebus},
where handshake-based synchronisation protocols and explicit timing
guards (e.g., the requirement that wired-OR signals persist longer
than $2 \times T_{p0}$ to be recognised as genuine) were adopted
precisely to avoid reliance on a shared global clock across
independently timed backplane modules.

The conventional taxonomy of clock relationships, following Teehan,
Greenstreet, and Lemieux~\cite{teehan2007}, distinguishes:

\begin{description}[style=nextline]
\item[Synchronous] Same frequency, known fixed phase.
\item[Mesochronous] Same frequency, unknown (possibly drifting) phase.
\item[Plesiochronous] Nearly the same frequency, slowly drifting phase.
\item[Heterochronous (Asynchronous)] Unrelated frequencies and phases.
\end{description}

A \textbf{bisynchronous FIFO} is a first-in-first-out buffer designed to
operate correctly across any of the latter three cases---that is, between
two independent clock domains with no guaranteed frequency or phase
relationship.  The word ``bisynchronous'' emphasizes that each side of the
FIFO is locally synchronous to its own clock, while the interface between
them is asynchronous.

What makes bisynchronous FIFOs remarkable, from the standpoint of the
FITO/Category Error debate~\cite{borrill2026whatwrong,borrill2026lamport},
is what they \emph{do not} require: they do
not require a global clock, a common timebase, synchronized timestamps,
or any form of time-of-day agreement between the two domains.  Instead,
they achieve correct, lossless, ordered data transfer using only
\textbf{interaction-based} primitives: handshake signals, pointer
comparisons, mutual exclusion elements, and causal flow-control protocols.

This paper surveys the bisynchronous FIFO literature and argues that it
constitutes a mature, silicon-proven existence proof against the FITO
assumption in distributed systems.  The argument proceeds in four stages:
we trace the historical development of bisynchronous designs
(Section~\ref{sec:history}), organize the literature into a taxonomy of
synchronization mechanisms (Section~\ref{sec:taxonomy}), draw out the
implications for the FITO/Category Error debate
(Section~\ref{sec:significance}), and identify open questions connecting
on-chip coordination to network-scale protocol design
(Section~\ref{sec:future}).

\section{Historical Development}
\label{sec:history}

\subsection{Precursors: FIFOs Across Timing Boundaries}

The idea that FIFO buffers could bridge independently timed domains
predates the GALS paradigm.  As early as 1974, Lim's NASA technical
report on ``quasi-perfect'' FIFOs~\cite{lim1974} explored iterative
FIFO implementations capable of operating in either synchronous or
asynchronous modes for the UNICON laser memory controller---an early
recognition that the boundary between clock domains is a design
problem requiring structural solutions rather than timing assumptions.

\subsection{IEEE P896 Futurebus (1979--1987): Handshake Protocols at the Backplane}

Concurrently with---and in some respects anticipating---the academic GALS
programme, the IEEE~P896 Futurebus working group (established 1979; active
design work from 1982) confronted the same synchronization problems at the
backplane level.  Borrill and Theus's 1984 paper in \emph{IEEE
Micro}~\cite{borrill1984futurebus} described an advanced communication
protocol for the proposed Futurebus that relied entirely on handshake-based
synchronisation rather than a shared bus clock.  The protocol defined a
taxonomy of fully interlocked handshake modes---single-source,
single-destination, source-and-destination (four-event and two-event
variants), and the TRI (three-party) handshake for multi-party
coordination---each guaranteeing correct data transfer across
independently timed backplane modules.

The Futurebus work is historically significant in several respects.
First, it predates Chapiro's 1984 GALS thesis (the working group had been
active since 1979 and Borrill's taxonomy was developed in the early
1980s).  Second, the protocol exploited both edges of the handshake
signal---a form of double-edge signalling that may represent one of the
earliest instances of what later became known as DDR (Double Data Rate)
in the memory industry.  Third, Borrill's subsequent doctoral
thesis~\cite{borrill1986futurebus}, which formalized the handshake taxonomy
and the associated transmission-line analysis, was later cited as prior art
in patent litigation challenging Rambus DDR memory patents---evidence of the
practical and legal weight carried by this early interaction-based design
work.

The IEEE~P896.1 standard was published in 1987 and established foundational
concepts in technology-independent bus synchronisation that would
influence subsequent interconnect standards throughout the late 1980s
and 1990s.

\subsection{Chapiro (1984): The GALS Thesis}

The Globally Asynchronous, Locally Synchronous (GALS) paradigm was
introduced by Daniel Chapiro in his 1984 Stanford doctoral
dissertation~\cite{chapiro1984}.  Chapiro recognized that forcing an
entire chip to share a single global clock was becoming untenable as
designs grew, and proposed instead that systems be composed of synchronous
islands communicating asynchronously.  He introduced the \emph{pausible
clocking} scheme, in which a local clock generator can be temporarily
halted to avoid metastability when an asynchronous input arrives at an
inconvenient moment.

Chapiro's insight was fundamentally architectural: the global clock is an
abstraction of convenience, not a physical necessity.  Each synchronous
island is a self-contained causal domain; what matters for correctness is
not that events share a common timestamp, but that data crosses boundaries
\emph{safely}---that is, without metastability and without data loss.
It is worth noting that the IEEE~P896 working group had independently
arrived at essentially the same conclusion at the backplane level several
years earlier, albeit from an engineering rather than a theoretical
starting point.

\subsection{Cummings (2001--2008): The Gray-Code Bisynchronous FIFO}

Clifford E.\ Cummings, in a series of influential SNUG (Synopsys Users
Group) papers~\cite{cummings2002,cummings2008}, established what became
the standard industrial approach to asynchronous FIFO design.  The key
contribution was the use of Gray-coded pointers---binary counters in which
only one bit changes per increment---synchronized across clock domains
using two-flip-flop (``brute force'') synchronizers.

The design works as follows.  A dual-port RAM stores the FIFO data.  The
write pointer is maintained in the write clock domain; the read pointer in
the read domain.  To compute the \texttt{full} flag, the read pointer must
be compared with the write pointer in the write domain, and vice versa
for the \texttt{empty} flag.  The Gray-coded pointers are passed through
two-stage synchronizer flip-flops into the opposite domain.  Because only
one bit changes per increment, any metastability event can at worst cause
the synchronized pointer to be one increment behind the true value---a
conservative error that causes the FIFO to appear slightly more full (or
more empty) than it actually is, but never causes overflow or underflow.

Cummings's papers are notable for their emphasis on formal reasoning about
correctness properties.  The key invariant is: the synchronized copy of a
pointer is always equal to or behind the true pointer.  Therefore, the
full flag may be asserted conservatively (when the FIFO is not yet truly
full) but never missed, and conversely for empty.  Put differently,
each side of the FIFO possesses \emph{boundedly stale knowledge} of
the remote pointer: not the instantaneous truth, but a delayed,
filtered observation that is guaranteed to be safe.  The full/empty
flags are epistemic statements---``given what I currently know after
synchronization latency, the FIFO is (not) full''---rather than
ontological claims about a shared-time global state.

This is a \textbf{causal safety} argument, not a timing argument.  The
correctness does not depend on the relative frequencies of the two clocks,
on any bound on clock skew, or on any timestamp comparison.  It depends
solely on the structural property of the Gray code and the metastability
resolution characteristics of the synchronizer flip-flops.

The penalty of brute-force synchronization is latency: each pointer
crossing requires at least two receiver clock cycles to resolve, adding
4+ cycles of round-trip synchronization delay.

\subsection{Chelcea and Nowick (2001--2004): Mixed-Timing Interfaces}

Tiberiu Chelcea and Steven M.\ Nowick at Columbia University developed a
family of low-latency mixed-timing FIFOs that could interface arbitrary
combinations of synchronous and asynchronous
systems~\cite{chelcea2001,chelcea2004}.  Their designs were implemented as
circular arrays of cells connected to common data buses, where data items
remain stationary once enqueued and are accessed via roving tokens.

Two key features of Chelcea and Nowick's work are particularly relevant.
First, their designs are \textbf{modular}: the put interface and get
interface at each cell are independent, and can be configured for either
synchronous (clocked) or asynchronous (handshake) operation.  This
modularity means the same FIFO cell can mediate sync-to-sync,
sync-to-async, async-to-sync, or async-to-async communication.

Second, they adapted Carloni et al.'s ``latency-insensitive''
protocol~\cite{carloni2001}---originally designed for single-clock systems
with long wires---to mixed-timing domains.  This migration demonstrated
that interaction-based flow control (valid/ready handshakes with
backpressure) generalizes naturally across clock domain boundaries.

The robustness guarantee is framed in terms of metastability probability:
the designs can be made ``arbitrarily robust'' by adding synchronizer
stages, trading latency for exponentially decreasing failure probability.

\subsection{Greenstreet and Chakraborty (1990--2003): STARI and Self-Timed Interfaces}

Mark Greenstreet's doctoral work introduced the STARI (Self-Timed At
Receiver's Input) protocol for mesochronous
communication~\cite{greenstreet1995}.  The key idea is to place a
self-timed FIFO at the boundary between two clock domains.  The FIFO
absorbs phase differences between the clocks: as long as the skew stays
within the FIFO's depth, data passes through without error.

Greenstreet \emph{proved} the correctness of STARI, including its
freedom from synchronization failures---a result that goes beyond
probabilistic arguments to a formal guarantee.  In subsequent work with
Chakraborty~\cite{chakraborty2003}, the approach was generalized to
single-stage FIFOs that exploit known clock relationships (rational
multiples, closely matched frequencies) to achieve synchronization with
minimal hardware.  This work received the Best Paper Award at ASYNC 2003.

The STARI philosophy is significant because it treats synchronization as
a \textbf{physical interaction} between the data stream and the local
clock, mediated by the elastic storage of the FIFO.  There is no reference
to wall-clock time, no timestamp, and no external coordination protocol.
The FIFO simply absorbs the mismatch, and the handshake protocol ensures
that data is never read before it is written and never overwritten before
it is read.

\subsection{Keller, Fojtik, and Khailany (2015): The NVIDIA Pausible Bisynchronous FIFO}

The 2015 NVIDIA paper, presented at the IEEE International Symposium on
Asynchronous Circuits and Systems~\cite{keller2015}, represents the state
of the art in bisynchronous FIFO design and is particularly important
because it was designed with industrial adoption in mind.

The design combines Chapiro's pausible clocking with Cummings's two-ported
synchronous FIFO structure.  Rather than synchronizing multi-bit Gray-coded
pointers through brute-force synchronizers, the pausible FIFO uses
\textbf{two-phase increment/acknowledge signals} to communicate pointer
updates between domains.  Each increment signal is synchronized through a
mutex-based pausible clock network that temporarily halts the receiver's
clock if the incoming signal arrives at an unsafe phase.

Key characteristics include an average latency of 1.34 clock cycles across
the asynchronous interface (compared to approximately 4 cycles for
brute-force synchronization); error-free operation via pausible clocking
that eliminates metastability rather than merely reducing its probability;
standard toolflow compatibility, since the FIFO memory is a conventional
dual-port synchronous RAM with only the mutex elements requiring custom
design; and scalable flow control through multiple increment/acknowledge
pairs that allow full throughput even when clock frequencies differ by up
to 2:1.

The flow-control protocol deserves detailed attention.  When the
transmitter writes a word to the FIFO, the write pointer logic toggles one
of the two-phase write pointer increment lines.  This toggle propagates
into the receiver's clock domain through the pausible synchronizer.  If the
toggle arrives at a safe clock phase, it passes through immediately.  If it
arrives at an unsafe phase, the mutex intervenes and the clock pauses until
the signal is safely resolved.  Once synchronized, the receiver's read
pointer logic updates its tracking of the write pointer and can proceed to
read.

The entire protocol is based on \textbf{causal handshakes}: the
transmitter cannot advance until it knows the receiver has acknowledged,
and the receiver cannot read until it knows the transmitter has written.
There is no global time, no timestamp, no clock synchronization protocol.
There are only interactions.

\subsection{Miro-Panades and Greiner (2007): The Bi-Synchronous FIFO for NoC}

Miro-Panades and Greiner~\cite{miropanades2007} contributed a
bi-synchronous FIFO design specifically targeted at Network-on-Chip (NoC)
architectures in GALS systems.  Their design introduced a novel encoding
algorithm for pointer comparison that avoids the overhead of Gray-code
conversion while maintaining the safety guarantee.

This work is significant because it explicitly positions the bisynchronous
FIFO as the fundamental building block for on-chip networking in GALS
architectures---a direct parallel to OAE's role as the fundamental
building block for inter-chip networking in multi-device
systems~\cite{borrill2026cap}.

\section{Taxonomy of Synchronization Approaches}
\label{sec:taxonomy}

The literature reveals a clear taxonomy of approaches to clock domain
crossing, organized by the mechanism used to achieve safety:

\begin{description}[style=nextline]
\item[Brute-Force Synchronization]
  Multi-flip-flop chains that allow metastability to resolve
  probabilistically.  High latency (2+ cycles per crossing), nonzero
  failure probability.  The Cummings FIFO~\cite{cummings2002}.

\item[Pausible Clocking]
  Mutex-based circuits that halt the local clock when an asynchronous
  input arrives at an unsafe phase.  Zero failure probability (the clock
  simply waits), very low average latency (approximately 1.3 cycles).
  The NVIDIA pausible bisynchronous FIFO~\cite{keller2015}.

\item[Self-Timed / Handshake-Based]
  Fully asynchronous FIFOs (micropipelines, GasP, Mousetrap) that use
  request/acknowledge handshakes for flow control.  No clock at all in
  the FIFO itself.  Sutherland's micropipelines~\cite{sutherland1989},
  Sutherland and Fairbanks's GasP~\cite{sutherland2001gasp}.

\item[Phase-Predictive]
  Circuits that exploit knowledge of clock phase relationships
  to sample at safe times.  Chakraborty and Greenstreet's rational-clock
  interfaces~\cite{chakraborty2003}.  Low latency, but require
  assumptions about clock stability.

\item[Mixed-Timing Modular]
  Chelcea and Nowick's family of FIFOs~\cite{chelcea2004}, which provide
  interchangeable synchronous and asynchronous interfaces.
  Latency-insensitive protocols adapted across timing domains.
\end{description}

What unites all of these approaches---and what distinguishes them from
the timestamp-based coordination used in distributed systems protocols
like PTP, TSN, and conventional Ethernet---is that \textbf{none of them
rely on a shared notion of time}.  They all achieve correctness through
structural and causal mechanisms: handshakes, mutual exclusion, pointer
invariants, and flow control.

\section{Significance for the FITO / Category Error Debate}
\label{sec:significance}

\subsection{The FITO Assumption in Distributed Systems}

The Forward-In-Time-Only (FITO) assumption, as articulated in the
DÆDÆLUS critique of conventional
networking~\cite{borrill2026whatwrong,borrill2026unix,borrill2026icloud},
holds that distributed systems protocols are designed around the premise
that events can be ordered by assigning timestamps from synchronized
clocks, and that these timestamps flow monotonically forward.  This
assumption manifests in PTP (IEEE~1588) and its variants, which attempt to
synchronize clocks across a network to nanosecond precision; TSN
(Time-Sensitive Networking), which schedules traffic based on a shared time
reference; conventional Ethernet's reliance on timeouts and retries for
error recovery; and high-frequency trading systems that depend on timestamp
ordering to determine trade priority, despite the fact that special
relativity makes simultaneous ordering of spatially separated events
physically meaningless~\cite{borrill2026hft}.

The ``category error''---in Ryle's original sense~\cite{ryle1949}, the
misapplication of concepts from one ontological category to
another---in the DÆDÆLUS critique is the confusion between
\emph{causal ordering} and \emph{temporal ordering}.  Lamport's 1978 paper
on logical clocks~\cite{lamport1978} already recognized that in a
distributed system, the only physically meaningful ordering is the
happened-before relation defined by message passing.  Yet the dominant
industrial practice remains to impose a global time coordinate and order
events by timestamp---an approach that is fragile (clocks drift), expensive
(synchronization protocols consume bandwidth and add latency), and in the
relativistic limit, physically
incoherent~\cite{borrill2026lamport,borrill2026cislunar}.

\subsection{Bisynchronous FIFOs as Existence Proof}

The bisynchronous FIFO literature provides a concrete, silicon-proven
existence proof that the category error is not merely a philosophical
objection but a practical one with a practical alternative.  Consider what
the NVIDIA pausible bisynchronous FIFO~\cite{keller2015} achieves:

\begin{enumerate}[nosep]
\item \textbf{Correct, lossless, ordered data transfer} between two
  completely independent clock domains.
\item \textbf{No shared timebase} of any kind---not even a common
  frequency.
\item \textbf{No timestamps}---pointer updates are communicated via
  interaction (two-phase handshakes through mutexes), not by encoding
  time values.
\item \textbf{Formal correctness guarantees} that are structural, not
  probabilistic (in the pausible case) or at worst probabilistically
  bounded with exponentially decreasing failure rates (in the
  brute-force case).
\item \textbf{Performance competitive with synchronous interfaces}---1.34
  cycles average latency, full throughput, lower energy than brute-force.
\end{enumerate}

If this can be achieved at the level of individual clock cycles on a
silicon die---the most demanding synchronization environment in all of
engineering---then the claim that distributed systems \emph{require}
global time synchronization to achieve reliable, ordered communication is
simply false.  The bisynchronous FIFO is a counterexample.

One might object that the FIFO itself embodies a forward-only
assumption: data enters at the write port and exits at the read port
in strict order, and this ordering is irreversible.  But this
confuses \emph{causal} ordering with \emph{temporal} ordering---precisely
the category error under discussion~\cite{ryle1949}.  The FIFO's
write-before-read invariant is a structural constraint: the write
pointer must advance before the read pointer can follow.  This is a
statement about \emph{what has happened} (a causal fact), not about
\emph{when it happened} (a temporal claim).  No timestamp is consulted;
the pointer comparison is an interaction-based check, not a
time-of-day comparison.  The FIFO's ordering is causal all the way
down.

\subsection{The Handshake as the Fundamental Primitive}

The deeper lesson of the bisynchronous FIFO literature is that the
\textbf{handshake}---not the timestamp---is the natural primitive for
coordination between independent causal domains.  Every bisynchronous
FIFO, regardless of its specific architecture, relies on some form of
handshake protocol:

\begin{itemize}[nosep]
\item In Cummings's design~\cite{cummings2002}, the Gray-coded pointer
  crossing through synchronizer flip-flops constitutes an implicit
  handshake: the pointer value is ``offered'' by the sender and
  ``accepted'' by the receiver after synchronization delay.
\item In the pausible designs~\cite{keller2015}, the two-phase
  increment/acknowledge protocol is an explicit handshake.
\item In Chelcea and Nowick's designs~\cite{chelcea2004}, put/get
  interfaces implement request/acknowledge protocols.
\item In Sutherland's micropipelines~\cite{sutherland1989}, the Muller
  C-element implements a rendezvous handshake between successive
  pipeline stages.
\item In STARI~\cite{greenstreet1995}, the self-timed FIFO mediates
  between the transmitter's data rate and the receiver's clock through
  elastic buffering---a physical handshake between signal propagation
  and clock edge.
\end{itemize}

This convergence is not accidental.  It reflects a fundamental physical
reality: two independent causal domains can only coordinate through
\emph{interaction}, not through shared observation of a clock.  The
handshake is the minimal interaction that establishes causal ordering
between sender and receiver.  It is, in the language of the OAE project,
an \emph{atomic swap}---an indivisible exchange of information that
creates a new causal fact~\cite{borrill2026cap,borrill2026flp}.

The same principle was established at the backplane level in the
IEEE~P896 Futurebus standard~\cite{borrill1986futurebus}, where
metastability and transmission-line phenomena (reflections, crosstalk,
and the wired-OR glitch) forced the adoption of interaction-based
synchronisation protocols.  Futurebus defined a taxonomy of
synchronous, asynchronous, and independent timing modes, and specified
handshake-based flow control across independently clocked backplane
modules---anticipating, in board-level engineering, exactly the design
philosophy that the GALS community would later formalize for on-chip
clock domain crossings.  The Futurebus experience demonstrates that
the handshake primitive scales across physical domains: from on-chip
flip-flop synchronizers, through backplane bus protocols, to
network-scale atomic swaps.

\begin{figure}[ht]
\centering
\includegraphics[width=\textwidth,trim=0 0 0 60,clip]{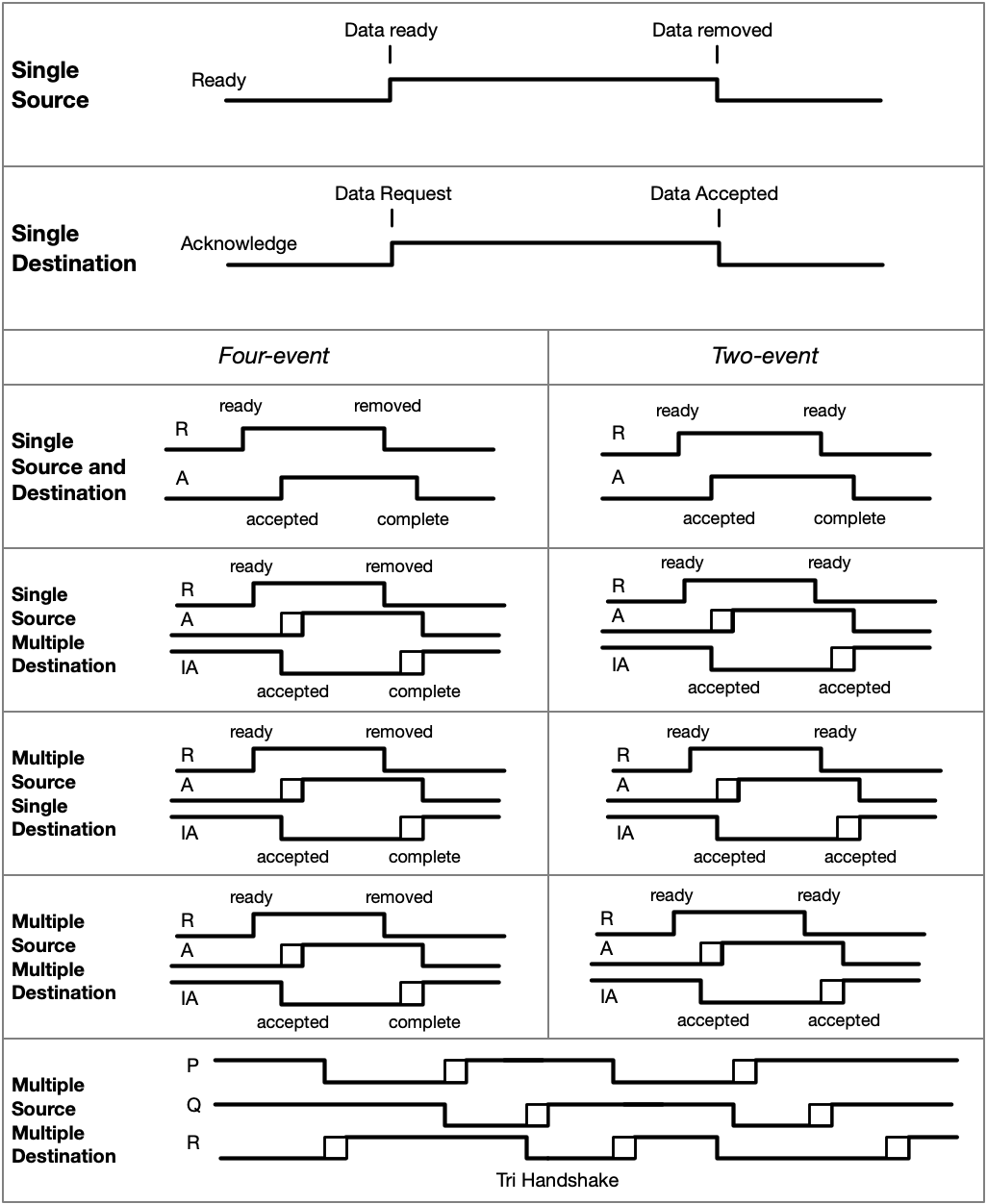}
\caption{Fully interlocked handshake protocols, from Borrill's 1986
  taxonomy~\cite{borrill1986futurebus}, originally developed for the
  IEEE~P896 Futurebus standard (working group est.\ 1979;
  see also~\cite{borrill1984futurebus}).  The figure classifies
  single-source, single-destination, source-and-destination (four-event
  and two-event variants), and the TRI handshake for multi-party
  coordination.  The rectangular pulse widths represent the wired-OR
  protocol window: the leading edge marks the moment the \emph{fastest}
  module asserts the wire, and the trailing edge marks the moment the
  \emph{slowest} module releases it.  This is fundamentally different
  from \emph{source-timed} protocols, where the transmitter holds a
  signal for a fixed, technology-dependent interval.  The wired-OR
  handshake is technology-independent by design: because the wire itself
  reports when all participants have responded, the protocol scales to
  any backplane length and any future silicon technology without
  modification.  This taxonomy predates by several years the
  formalizations of four-phase and two-phase handshakes in the
  asynchronous VLSI literature (Sutherland 1989, Chelcea and
  Nowick 2001).}
\label{fig:handshake-taxonomy}
\end{figure}

\subsection{From On-Chip to Inter-Chip: The OAE Parallel}

The bisynchronous FIFO mediates between two clock domains on a chip.
The IEEE~P896 Futurebus~\cite{borrill1986futurebus} mediated between
independently timed modules on a backplane.  OAE mediates between
nodes on a network.  The structural parallel spans three physical
scales:

\begin{center}
\begin{tabular}{llll}
\toprule
\textbf{Concept} & \textbf{Bisync FIFO} & \textbf{Futurebus} & \textbf{OAE} \\
\midrule
Independent domains & Clock domains & Backplane modules & Network nodes \\
Ordering mechanism & Pointer handshakes & Bus handshakes & Atomic swaps \\
Flow control & Full/empty flags & Wired-OR protocol & Credit-based backpressure \\
Data integrity & Gray-code invariants & $2 \times T_{p0}$ guard & ACID transactions \\
Global time required & No & No & No \\
Timeout/retry needed & No & No & No \\
\bottomrule
\end{tabular}
\end{center}

The GALS literature's trajectory---from Chapiro's observation that global
clocks are unnecessary~\cite{chapiro1984}, through decades of refinement
of interaction-based synchronization, to silicon-proven designs that
outperform their synchronous counterparts---prefigures the trajectory that
OAE proposes for networking: from the observation that global time
synchronization is a category
error~\cite{borrill2026whatwrong,borrill2026lamport}, through the
development of interaction-based networking primitives, to a protocol stack
that achieves atomicity without timestamps.

There is, however, an important sense in which the GALS literature stops
short of the strongest coordination primitive.  Chelcea and Nowick's
``put/get'' interface~\cite{chelcea2004}---and indeed every bisynchronous
FIFO---decomposes the exchange into two temporally ordered half-operations:
the producer \emph{puts} data into the buffer, and at some later time the
consumer \emph{gets} it.  Between the put and the get there exists an
intermediate state in which the data has been committed by the sender but
not yet consumed by the receiver.  The full/empty flags manage this
window, but they do not eliminate it.  The OAE atomic
swap~\cite{borrill2026cap,borrill2026flp} collapses the two half-operations
into a single indivisible causal event: data transfer and acknowledgement
occur as one atomic action, with no intermediate state visible to either
party.  Put differently, the bisynchronous FIFO proves that you do not
need \emph{time} to coordinate between independent domains; OAE goes
further by showing that you do not even need the \emph{two-phase
decomposition} into separate write and read operations.  The atomic swap
is the natural endpoint of the trajectory that the GALS literature
initiated.

\subsection{Lamport's Logical Clocks: The Bridge}

It is worth noting that Lamport's 1978 paper~\cite{lamport1978} occupies
a curious intermediate position.  Lamport recognized that the only
physically meaningful ordering in a distributed system is the
happened-before relation defined by message passing---a fundamentally
interaction-based insight.  Yet his paper also introduced the idea of
using logical \emph{timestamps} (monotonic counters) to represent this
ordering, and further proposed physical clock synchronization to handle
``anomalous'' cases where the logical ordering diverges from the
physical ordering perceived by external
observers~\cite{borrill2026lamport}.

The bisynchronous FIFO literature resolves this tension by showing that
you do not need even logical timestamps to achieve correct ordering.  A
pointer is not a timestamp---it is a \emph{position} in a causal
sequence.  The write pointer says ``this is the next slot to write''; the
read pointer says ``this is the next slot to read.''  Their comparison
determines whether the FIFO is full or empty.  No notion of ``when''
enters the picture at all.  Only ``what has been done'' and ``what
remains to be done.''

This is the sense in which the bisynchronous FIFO embodies an
\emph{interaction-based} approach to coordination: the relevant
information is not ``what time is it?'' but ``have we interacted?''  The
pointer handshake is a physical record of causal connection, not a
temporal assertion.

\subsection{Implications for High-Frequency Trading}

The bisynchronous FIFO literature also sharpens the HFT
critique~\cite{borrill2026hft}.  If two exchanges operate as independent
clock domains (which they physically are, separated by lightspeed delays),
then ordering trades by timestamp is precisely the category error that
bisynchronous FIFO designers learned to avoid decades ago.  The correct
approach is to order trades by \emph{causal interaction}---which trade was
actually received and acknowledged first by the matching engine, regardless
of what timestamp the sender attached.  The matching engine's receive
buffer is, in essence, a bisynchronous FIFO: data arrives from an external
clock domain (the trader's system) and must be safely ingested into the
engine's local domain.  The ordering is determined by arrival order in the
FIFO, not by any timestamp.

\section{Open Questions and Future Work}
\label{sec:future}

Several questions remain open at the intersection of bisynchronous FIFO
design and the FITO/Category Error debate:

\paragraph{Scaling to multi-hop networks.}
On-chip GALS systems communicate between adjacent clock domains.  OAE
must handle multi-hop communication where the number of independent
domains scales with network size.  The NoC literature~\cite{miropanades2007,dimitrakopoulos2016} addresses this for
on-chip networks; extending these results to datacenter-scale fabrics is
the central challenge.

\paragraph{Formal verification at the protocol level.}
The bisynchronous FIFO literature has strong formal verification results
at the circuit level (Greenstreet's correctness
proof~\cite{greenstreet1995}).  Extending these methods to the OAE
protocol stack---proving that the interaction-based primitives maintain
ACID properties across a network of bisynchronous links---would be a
significant contribution.

\paragraph{Bridging the SerDes gap.}
Modern high-speed serial links (PCIe, Ethernet SerDes) already perform
clock-domain crossing at the physical layer, often using elastic
buffers that are functionally bisynchronous FIFOs.  Making this
implicit FIFO behavior explicit and accessible to higher protocol
layers---rather than hiding it behind a synchronous abstraction---is
part of the OAE SerDes layer specification.

\paragraph{Category-theoretic formalization.}
The structural parallel between bisynchronous FIFOs and OAE atomic
swaps suggests that both may be instances of a more general
category-theoretic construction (perhaps related to Pratt's
Pomsets or to monoidal categories of concurrent processes).
Formalizing this connection could provide a unified framework for
reasoning about interaction-based coordination at all scales.

\paragraph{Connection to message passing without temporal direction.}
The constraint-semantics approach to message
passing~\cite{borrill2026msgpassing} provides a formal framework in
which messages are boundary constraints rather than temporal events.
The bisynchronous FIFO's pointer handshake is a physical instantiation
of this constraint semantics: the write pointer constrains what the
reader may access, and the read pointer constrains what the writer may
overwrite, without either side referencing time.

\section{Conclusion}
\label{sec:conclusion}

The bisynchronous FIFO is one of the most thoroughly studied and
practically validated structures in digital engineering.  Over four decades,
the GALS community has developed, formally verified, and commercially
deployed designs that achieve correct, high-performance data transfer
between fully independent clock domains using only interaction-based
primitives---handshakes, mutual exclusion, pointer invariants, and causal
flow control.  No global time, no timestamps, no clock synchronization
protocol is required.

This body of work constitutes a powerful existence proof for the
DÆDÆLUS critique of conventional distributed
systems~\cite{borrill2026whatwrong}: the reliance on global time
synchronization in distributed systems protocols is a category error, and
interaction-based primitives can replace timestamp-based coordination
without sacrificing correctness or performance.  The bisynchronous FIFO
shows, in proven silicon, that the right way to coordinate between
independent causal domains is not to synchronize their clocks but to
synchronize their \emph{interactions}.  Open Atomic
Ethernet~\cite{borrill2026cap,borrill2026flp} proposes to apply this
lesson---learned in the most demanding synchronization environment in
engineering---to the design of network protocols.

\section*{Acknowledgments}

This paper was developed with the assistance of AI tools (Claude,
Anthropic) for literature survey organization, drafting, and LaTeX
preparation.  The research program and all technical claims are the sole
responsibility of the author.  The FITO category mistake framework
originates from the author's work on hardware-verified deterministic
networking at Earth Computing (2012--2020) and continues under the
DÆDÆLUS research program.


\end{document}